\documentclass[traditabstract,preprint2]{aa}
\usepackage{txfonts}
\usepackage{indentfirst}
\usepackage{amssymb}
\usepackage{epsfig, amsmath, amsfonts}
\usepackage{natbib}
\bibpunct{(}{)}{;}{a}{}{,}

\usepackage{epstopdf}

\usepackage{verbatim}

\begin{document}


\title{On beryllium-10 production in gaseous protoplanetary disks and implications on the astrophysical setting of refractory inclusions}

\author{Emmanuel Jacquet}

\institute{Mus\'{e}um national d'Histoire naturelle, IMPMC, D\'{e}partement Origines et Evolution, 57 rue Cuvier, 75005 Paris, France \email{emmanuel.jacquet@mnhn.fr}}


\keywords{accretion, accretion disks -- Sun: flares -- Stars: protostars --  cosmic rays -- meteorites, meteors, meteoroids -- X-rays: stars}

\abstract{Calcium-Aluminum-rich Inclusions (CAIs), the oldest known solids of the solar system, show evidence for the past presence of short-lived radionuclide beryllium-10, which was likely produced by spallation during protosolar flares. While such $^{10}$Be production has hitherto been modeled at the inner edge of the protoplanetary disk, I calculate here that spallation at the disk surface may reproduce the measured $^{10}$Be/$^9$Be ratios at larger heliocentric distances. Beryllium-10 production in the gas prior to CAI formation would dominate that in the solid. Interestingly, provided the Sun's proton to X-ray output ratio does not decrease strongly, $^{10}$Be/$^9$Be at the CAI condensation front would increase with time, explaining the reduced values in a (presumably early) generation of CAIs with nucleosynthetic anomalies. CAIs thus need not have formed very close to the Sun and may have condensed at 0.1-1 AU where sufficiently high temperatures originally prevailed.}   

\titlerunning{$^{10}$Be production in disk gas}
\authorrunning{E. Jacquet}

\maketitle

\section{Introduction}

  Primitive meteorites, or \textit{chondrites}, are conglomerates of solids which formed and accreted in the solar protoplanetary disk. Among those, the oldest are the \textit{refractory inclusions}, comprising \textit{Calcium-Aluminium-rich Inclusions} (CAIs) and \textit{Amoeboid Olivine Aggregates} (AOA) \citep[e.g.][]{MacPherson2014,Krotetal2004}. Thermodynamic calculations predict that CAIs should be the first condensates in a cooling gas of solar composition \citep{Grossman2010,DavisRichter2014}. Since the required temperatures would be in the range 1500-2000 K \citep{Woodenetal2007}, it is widely assumed that refractory inclusions formed close to the Sun, but precisely how close and in what astrophysical setting remains unclear \citep{Wood2004, Jacquet2014review}.

  An important clue in this respect may be provided by beryllium-10, a short-lived radio-isotope decaying into boron-10 with a half-life of 1.5 Ma, and which, unlike other known extinct radionuclides such as aluminum-26, is not produced by stellar nucleosynthesis but may be formed through spallation by energetic particles \citep{DavisMcKeegan2014,Lugaroetal2018}. Indeed, since the original discovery by \citet{McKeeganetal2000}, all CAIs with analytically suitable Be/B have shown $^{10}$B/$^{11}$B excesses correlated therewith. The slopes of the resulting isochrons translate into initial (i.e. upon the last equilibration) $^{10}$Be/$^{9}$Be ratios averaging $\sim 6\times 10^{-4}$ in CV chondrite (type A and B) CAIs \citep{DavisMcKeegan2014}, with less than a factor of two spread. Yet systematically lower values around $3\times 10^{-4}$ and $5\times 10^{-4}$  have been found for two CV chondrite FUN CAIs ("Fractionated and Unknown Nuclear effects"; \citet{MacPhersonetal2003,Wielandtetal2012}) and platy hibonite crystals (PLAC) in CM chondrites \citep{Liuetal2009,Liuetal2010}, respectively, and, conversely, values up to $10^{-2}$ have been found for Isheyevo CAI 411 \citep{Gounelleetal2013} and more recently, fine-grained group II CV chondrite CAIs \citep{Sossietal2017}. The overall spread, along with correlated $^{50}$V excesses (also ascribed to spallation) in the latter objects \citep{Sossietal2017}, is difficult to reconcile with simple $^{10}$Be inheritance from the (galactic cosmic ray-irradiated) protosolar cloud \citep{Deschetal2004}, which should be largely homogeneous, and argues in favor of local production in the disk following flares from the young Sun \citep{Gounelleetal2001,Gounelleetal2006,Gounelleetal2013,Sossietal2017}, such as those manifested in X-rays by present-day protostars \citep[e.g.][]{Feigelsonetal2002,Wolketal2005, Preibischetal2005, Telleschietal2007, Guedeletal2007, Bustamanteetal2016} or evidence for enhanced ionization in some protostellar envelopes \citep{Ceccarellietal2014,Favreetal2017, Favreetal2018}.
  
  Since the energetic protons would not penetrate further than $\sim 690\:\mathrm{kg/m^2}$ \citep{UmebayashiNakano1981} in the gas, \citet{Sossietal2017} suggested that CAIs spent a few thousand orbits at the inner edge of the protoplanetary disk, similar to earlier studies which had adopted the framework of the X-wind scenario \citep{Leeetal1998, Shuetal2001, Gounelleetal2001}. However, among the objections to the latter discussed by \citet{Deschetal2010}, a general issue is that CAI-forming temperatures would be expected significantly further outward (0.1-1 AU), because of local dissipation of turbulence. If nonetheless CAIs 
 \textit{all} went somehow through this narrow region, escaped accretion to the Sun or ejection from the solar system 
they would be essentially "lucky" foreign material in the chondrites which incorporated them. Yet CAIs are present in carbonaceous chondrites at about the abundances predicted by \textit{in situ} condensation \citep{Jacquetetal2012S}. Also, although (non-CI) carbonaceous chondrites are enriched in refractory elements relative to CI chondrites, this cannot be ascribed to simple CAI addition to CAI-free CI chondritic material for should the CAIs be mentally subtracted, these chondrites would have subsolar refractory element abundances \citep{Hezeletal2008}, not even considering those CAIs that were converted into chondrules \citep{MisawaNakamura1988,JonesSchilk2009, MetzlerPack2016, EbertBischoff2016, JacquetMarrocchi2017,Marrocchietal2018}. Isotopic systematics for elements of different volatilities also indicate a need for a non-refractory isotopically CAI-like component in chondrites \citep[e.g.][]{Nanneetal2019}
. So there is evidence for a genetic relationship between CAIs and (part of) their host carbonaceous chondrites which argues against an origin at the disk inner edge, whose contribution to distant chondritic matter would likely be minor.	
	
\begin{figure}
\resizebox{\hsize}{!}{
\includegraphics{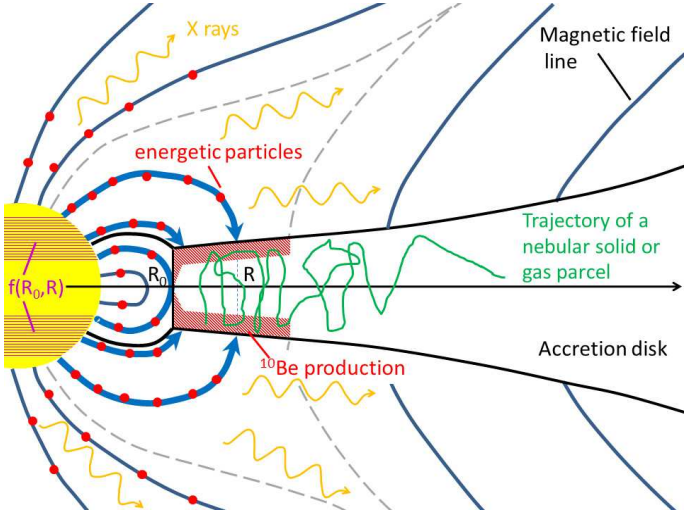}
}
\caption{Sketch of the scenario of $^{10}$Be production developed in this paper. Solar cosmic rays cause spallation reactions in the upper layers of the disk, in the gas and/or condensates, within a region (outlined by a dashed separatrix) where magnetic field lines originating from the proto-Sun and its outskirts hit the disk. The field and particle emission geometries are purely illustrative and do not detract the generality of the formalism.}
\label{sketch}
\end{figure}

	However, the inner edge of the disk is not the only region where CAIs or their precursors could have been exposed to energetic solar protons. CAIs floating further out in the disk would at times reach the upper layers of the disk. While such excursions may individually incur modest proton fluences, their cumulative contributions might explain the total $^{10}$Be evidenced in CAIs. Moreover, prior to CAI condensation, the gas exposed at the surface of the disk may also undergo spallation and pass on the then-produced $^{10}$Be to the later-formed condensates.  The purpose of this paper is to analytically calculate the amount of beryllium-10 produced by such channels and thence evaluate the viability of having CAIs form at 0.1-1 AU from the young Sun against their irradiation record. Figure \ref{sketch} provides a sketch of the overall scenario and setting. After several generalities in Section \ref{Generalities}, I will express the $^{10}$Be/$^9$Be ratio in the disk and free-floating solid CAIs in Section \ref{Results}. I then discuss the numerical evaluations thereof and their implications in Section \ref{Discussion} before concluding in Section \ref{Conclusion}.

\section{Prolegomena}
\label{Generalities}

\subsection{Disk model}

I consider a protoplanetary disk in a cylindrical coordinate system 
 with heliocentric distance $R$. The inner regions can be described under the steady state approximation so long the evolution timescales of the system are long compared to their local viscous timescale
\begin{eqnarray}
t_{\rm vis}\left(R\right) &\equiv & \frac{R^2}{\nu}\nonumber\\
&=& 0.02\:\mathrm{Ma}\left(\frac{R}{0.5\:\rm UA}\right)^{1/2}\left(\frac{1500\:\rm K}{T}\right)\left(\frac{10^{-3}}{\alpha}\right),
\end{eqnarray}
where $\nu=\alpha c_s^2/\Omega_K$ is the effective turbulent viscosity, $\alpha$ the dimensionless turbulence parameter \citep[e.g.][]{Jacquet2013}, $c_s=\sqrt{k_BT/m}$ the isothermal sound speed with $k_B$ the Boltzmann constant, $T$ the temperature and $m=3.9\times 10^{-27}\:\rm kg$ the mean molecular mass, and $\Omega_K$ the Keplerian angular velocity. Since $t_{\rm vis}(R)$ is also much shorter than the half-life of $^{10}$Be, I will neglect its decay during the CAI formation epoch. This is consistent with the short timescales (no more than a few hundreds of millenia) of CAI formation and (isotopically resetting) re-heating events \citep[e.g.][]{MacPherson2014}, which should have ended after the last production of $^{10}$Be so as to account for the isochron behavior of the Be-B system in each CAI (that is, with $^{10}$B excesses scaling with Be rather than the target nuclides leading to its parent; see next subsection).

  If infall from the parental cloud can be neglected \textit{in the inner region}, the disk mass accretion rate $\dot{M}=-2\pi R\Sigma u_R$ (with $\Sigma$ the disk surface density and $u_R$ the net, turbulence-averaged, gas radial velocity) is uniform and obeys:
\begin{equation}
\label{steady standard}
\dot{M}\left(1-\sqrt{\frac{R_0}{R}}\right)=3\pi\Sigma\nu,
\end{equation}
or, equivalently,
\begin{equation}
\label{uR}
u_R=-\frac{3\nu}{2R\left(1-\sqrt{R_0/R}\right)},
\end{equation}
where we have assumed that the stress vanishes at the disk inner edge $R_0$ \citep[e.g.][]{BalbusHawley1998}. When calculating $^{10}$Be/$^9$Be in later sections, I shall show expressions for a general $\dot{M}(R)$ before specializing to the case of uniform mass accretion rate. Appendix \ref{Infall} explores the corrections of infall to equation (\ref{steady standard}).

 If I inject the temperature due to viscous dissipation of turbulence (e.g. appendix A of \citet{Jacquetetal2012S}) in equation (\ref{steady standard}), I obtain:
\begin{eqnarray}
T &=& \left(\frac{3\kappa m \dot{M}^2\left(1-\sqrt{R_0/R}\right)^2\Omega_K^3}{128\pi^2\sigma_{SB}k_B\alpha}\right)^{1/5}\nonumber\\
&=& 1200\:\mathrm{K} \left(\frac{\kappa}{10^{-2}\:\rm m^2/kg}\right)^{1/5}\left(\frac{10^{-3}}{\alpha}\right)^{1/5}\left(1-\sqrt{\frac{R_0}{R}}\right)^{2/5}\nonumber\\
&& \left(\frac{0.5\:\rm AU}{R}\right)^{9/10}\left(\frac{\dot{M}}{10^{-7}\:\rm M_\odot/a}\right)^{2/5}
\label{T}
\end{eqnarray}
with $\kappa$ the specific Rosseland mean opacity and $\sigma_{SB}$ the Stefan-Boltzmann constant. This confirms that CAI-forming temperatures may be reached at a fraction of an AU for $\dot{M}\gtrsim 10^{-7}\:\rm M_\odot/a$.

\subsection{$^{10}$Be production rate}
\label{production rate}

The local production rate of $^{10}$Be by spallation normalized to $^9$Be may be written as \citep[e.g.][]{Sossietal2017}: 
\begin{equation}
\label{local 10Be}
\left(\frac{\overset{\bullet}{^{10}\rm Be}}{^9\rm Be}\right)_{\rm tg}=\int_0^{+\infty}\mathrm{d}E\sigma_{ki}(E)\left(\frac{k}{\rm Be}\right)_{\rm tg} \int\mathrm{d}\Omega I(i,E,\Omega)
\end{equation}
with $\sigma_{ki}$ the $^{10}$Be production cross section for target (tg) nuclide $k$ and cosmic ray species $i$ (with energy per nucleon $E$) and $I(i,E,\Omega)$ the corresponding (orientation-dependent) monoenergetic cosmic ray specific intensity (defined by number and not energy, as in \citealt{Gounelleetal2001}). Here and throughout, all isotopic or elemental ratios are atomic and the Einstein convention for summation over the repeated indices $k$ and $i$ is adopted. I assume that the incoming cosmic rays have a solar composition (but see \citet{Mewaldtetal2007} for details on contemporaneous solar energetic particles), with e.g. $^4$He/H=0.1, and obey a power law $\propto E^{-p}$ for $E\geq E_{\rm 10}\equiv 10\:\rm MeV$ upon arrival on the disk. I will ignore the contribution of secondary neutrons, despite their comparable cross sections \citep[e.g.][]{LeyaMasarik2009} and larger attenuation columns. This is because a dilute medium such as the surface of the disk will let free decay thwart significant accumulation of neutron flux \citep{UmebayashiNakano1981}. However, based on this latter work, this is only marginally true (see equation (\ref{rho_spall}) in appendix \ref{Settling}), so the production rates given here should be strictly viewed as a \textit{lower} bounds. 

  In the following sections, we will be interested in the density ($\rho$)-weighted vertical average of the production rate at a given heliocentric distance:
	\begin{equation}
	\left\langle\left(\frac{\overset{\bullet}{^{10}\rm Be}}{^9\rm Be}\right)_{\rm tg}\right\rangle_{\rho}\equiv\frac{1}{\Sigma}\int_0^{\Sigma}\mathrm{d}\Sigma'\left(\frac{\overset{\bullet}{^{10}\rm Be}}{^9\rm Be}\right)_{\rm tg}\approx \frac{2}{\Sigma}\int_0^{+\infty}\mathrm{d}\Sigma'\left(\frac{\overset{\bullet}{^{10}\rm Be}}{^9\rm Be}\right)_{\rm tg,+}
	\end{equation}
where $\Sigma'$ is the column density integrated vertically from the upper surface and the $+$ subscript in the final equality denotes restriction to cosmic rays from the upper side\footnote{Assuming symmetry of irradiation about the midplane, although this is not a prerequisite to the final result (equation (\ref{avg mit Kp})).}, assuming $\Sigma$ is much larger than the attenuation column $\Sigma_{p,i}(E)$ of the cosmic rays, defined here as:
\begin{equation}
\Sigma_{p,i}(E)\equiv\frac{1}{I_+(i,E,\Omega;0)}\int_0^{+\infty}\mathrm{d}\Sigma'' I_+(i,E,\Omega;\Sigma'') 
\end{equation} 
where $\Sigma''$ is the actually traversed column density. If I envision a cosmic ray beam (with the particles gyrating along a magnetic field line) penetrating the disk with a grazing angle $\phi$ and ignore secondary particles, $\Sigma''=\Sigma'/(\mu\mathrm{sin}\phi)$ with $\mu$ the cosine of the pitch angle of the cosmic ray with respect to the magnetic field line \citep{Padovanietal2018}. I thus obtain:
\begin{equation}
\label{avg mit Sigmai}
\left\langle\left(\frac{\overset{\bullet}{^{10}\rm Be}}{^9\rm Be}\right)_{\rm tg}\right\rangle_{\rho}=\frac{2}{\Sigma}\int_0^{+\infty}\mathrm{d}E\sigma_{ki}(E)\Sigma_{p,i}(E)\left(\frac{k}{\rm Be}\right)_{\rm tg}\left(\frac{i}{\rm H}\right)_{\rm CR}\frac{\mathrm{d}\mathbf{F}_{H}}{\mathrm{d}E}\cdot \mathbf{n}
\end{equation}  
where $\mathbf{n}$ is the downward-directed unit vector normal to the disk surface and $\mathrm{d}\mathbf{F}_H/\mathrm{d}E$ the incoming differential proton number vector flux. If I scale the latter to the incoming $>$ 10 MeV/nucleon energy vector flux $\mathbf{F}_{10}$ and extract O as a proxy for all $^{10}$Be-producing target nuclides, equation (\ref{avg mit Sigmai}) can be further manipulated to finally yield:
\begin{equation}
\label{avg mit Kp}
\left\langle\left(\frac{\overset{\bullet}{^{10}\rm Be}}{^9\rm Be}\right)_{\rm tg}\right\rangle_{\rho}=\frac{K_pL_{10}}{2\pi R\Sigma}\left(\frac{\rm O}{\rm Be}\right)_{\rm tg}\frac{\partial f(R_0,R)}{\partial R}
\end{equation}
with
\begin{equation}
\label{Kp}
K_p\equiv \left(p-2\right)E_{10}^{p-2}\left(\frac{k}{\rm O}\right)_{\rm tg}\left(\frac{i}{\rm H}\right)_{\rm CR}\int_0^{+\infty}\mathrm{d}E \sigma_{ki}(E)\Sigma_{p,i}(E)E^{-p},
\end{equation}
which is plotted in Fig. \ref{Kp plot} (see Appendix \ref{Kp calculation} for further details on its calculation and data sources). $f(a,b)$ is the fraction of the total $>$10 MeV/nucleon energy luminosity $L_{10}$ which reaches the disk between heliocentric distances $a$ and $b$. The latter depends on the geometry of energetic particle emission and the magnetic field configuration around the young Sun, which are largely unknown (protosolar corona, star-disk fields or disk fields; \citet{Feigelsonetal2002,DonatiLangstreet2009}). For reference, in the case of a geometrically flared disk, an isotropic point source with straight trajectories would yield $f(R_0,R)=H_{\rm spall}/R$ with $H_{\rm spall}$ the height of the spallation layer above the midplane (see Appendix \ref{Settling} for an estimate). For illustration in the plots of the next section, I will use a form of $f$ assuming that the outgoing energy flux density of the energetic particles is uniform over the Sun-centered sphere of radius $R_0$ and that the magnetic field is approximately dipolar (with a magnetic moment along the rotation axis of the disk) for fieldlines interior to a maximum irradiation distance $R_{\rm max}$. This yields\footnote{If we follow a field line from $(r,\theta)=(R_0,\theta_0)$ in polar coordinates, we have 
\begin{equation}
\frac{\mathrm{d}r}{r\mathrm{d}\theta}=\frac{B_r}{B_\theta}=\frac{2\mathrm{cos}\theta}{\mathrm{sin}\theta},
\end{equation}
so that upon crossing the midplane $r=R=R_0/\mathrm{sin}^2\theta_0$ with $f(R_0,R)=\mathrm{cos}\theta_0$.
}:
\begin{equation}
f(R_0,R)=\mathrm{max}\left(\sqrt{1-\frac{R_0}{R}},\sqrt{1-\frac{R_0}{R_{\rm max}}}\right).\nonumber
\end{equation}  
However, in the main text, I will keep general expressions as functions of $f$. This parameterization of our ignorance will turn out to be quite handy in the calculation. 

\begin{figure}
\resizebox{\hsize}{!}{
\includegraphics{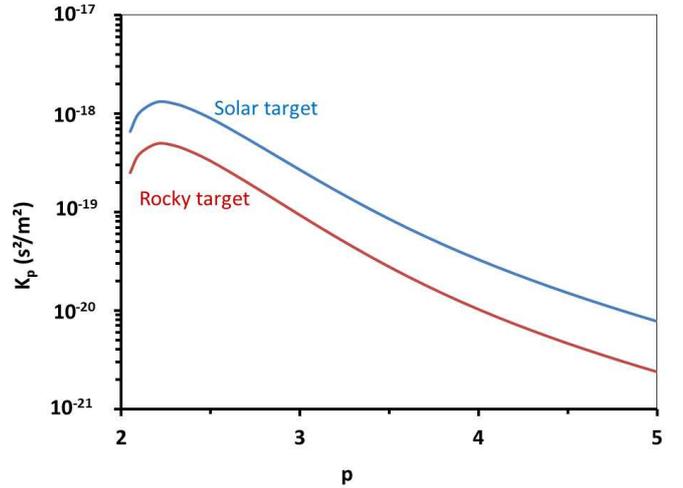}
}
\caption{Plot of $K_p$ as a function of the power law exponent of the incoming cosmic ray energy distribution, for a solar (without possible implantation contributions) and a rocky target (see subsections \ref{Gas} and \ref{Solids}, respectively).}
\label{Kp plot}
\end{figure}

  A means of comparison with previous models invoking irradiation of bare solids \citep[e.g.][]{Leeetal1998,Gounelleetal2006,Sossietal2017} is to divide the rate given by equation (\ref{avg mit Kp}) by that in a target of the same composition exposed to the same unattenuated proton flux. Assuming isotropic distribution of cosmic rays (in each half-space along the field line), this gives:
\begin{equation}
\frac{\overset{\bullet}{^{10}\rm Be}}{\overset{\bullet}{^{10}\rm Be}_{\rm bare}}=\frac{\langle \Sigma_p(E)\rangle_{\sigma E^{-p}}\mathrm{sin}\phi}{\Sigma}
\end{equation}
  with 
\begin{eqnarray}
\langle \Sigma_p(E)\rangle_{\sigma E^{-p}}\equiv \left(\frac{k}{\rm O}\right)_{\rm tg}\left(\frac{i}{\rm H}\right)_{\rm CR}\int_0^{+\infty}\mathrm{d}E \sigma_{ki}(E)\Sigma_{p,i}(E)E^{-p}\nonumber\\
\Bigg/\left[\left(\frac{k}{\rm O}\right)_{\rm tg}\left(\frac{i}{\rm H}\right)_{\rm CR}\int_0^{+\infty}\mathrm{d}E \sigma_{ki}(E)E^{-p}\right],
\end{eqnarray}
which (for a refractory target) is 90 kg/m$^2$ for p = 2.5 typical of gradual flares favored by \citet{Gounelleetal2013} and \citet{Sossietal2017}, and 5 kg/m$^2$ for p = 4 commensurate with impulsive flares (with subdominant fluences in the present-day Sun; \citealt{DesaiGiacalone2016}). Obviously, for $\Sigma\gg 10^4\:\rm kg/m^2$ (the Minimum Mass Solar Nebula at 1 AU; \citealt{Hayashi1981}) this calls for (intermittent) irradiation timescales $2\pm 1$ orders of magnitude above the centuries calculated by \citet{Sossietal2017}. As alluded in the introduction, the transport timescale $t_{\rm vis}$ may however provide the correct order of magnitude. The following section undertakes to calculate more precisely the net $^{10}$Be abundances produced in gas and  solids and during radial transport.

\section{$^{10}$Be abundances}
\label{Results}

\subsection{Bulk $^{10}$Be/$^9$Be in the disk}
\label{Gas}

  In this subsection, I calculate the average $^{10}$Be/$^9$Be of the disk as a function of heliocentric distance, irrespective of its (temperature-dependent) physical state (condensation fraction) to which spallation reactions are insensitive. I assume solar abundances throughout the inner disk owing to tight coupling of the condensates with the gas \citep[e.g.][]{Jacquetetal2012S}. 

  A steady-state gradient of $^{10}$Be/$^9$Be should arise in the disk because production of $^{10}$Be is balanced by loss to the Sun by accretion. The transport equation reads:
\begin{eqnarray}
\label{transport 10Be}
\frac{1}{R}\frac{\partial}{\partial R}\Bigg[R\Bigg(\Sigma u_R c_{\rm Be}\left(\frac{^{10}\rm Be}{^9\rm Be}\right)_{\rm disk}
-D_R\Sigma\frac{\partial}{\partial R}\left(c_{\rm Be}\left(\frac{^{10}\rm Be}{^9\mathrm{Be}}\right)_{\rm disk}\right)\Bigg)\Bigg]\nonumber\\
=\Sigma c_{\rm Be}\left\langle\left(\frac{\overset{\bullet}{^{10}\rm Be}}{^9\rm Be}\right)_{\rm disk}\right\rangle_{\rho}
= \frac{K_{p,\rm disk}L_{10}c_{\rm O}}{2\pi R}\frac{\partial f(R_0,R)}{\partial R}
\end{eqnarray} 
with $c_{\rm Be, O}$ the Be, O concentration in number per unit mass of solar gas and $D_R$ the turbulent diffusion coefficient. The $K_{p,\rm disk}$ factor here includes contributions from $k=^{12}$C, $^{16}$O and $^{14}$N (even though the last one is unimportant, with N/O=0.14, compared to C/O=0.5; \citealt{Lodders2003}) and $i$=$^1$H, $^4$He. In addition, it includes contributions from \textit{indirect reactions} \citep{Deschetal2004}, that is those where the heavy nuclei originate from the cosmic rays instead of the target, since, although the hereby produced $^{10}$Be will retain the momentum of the incoming particles, it should be stopped further downstream in the disk. Since Galilean invariance mandates $\sigma_{ik}=\sigma_{ki}$, this amounts to multiplying each "direct" reaction contribution in the right-hand-side of equation (\ref{Kp}) by $1+\left[(k/i)_{\rm CR}/(k/i)_{\rm disk}\right]\Sigma_{p,k}(E)/\Sigma_{p,i}(E)=1+A_k/Z_k^2$ where the second expression uses the assumption that the disk has a (solar) composition, identical to the cosmic rays' (see also Appendix \ref{Kp calculation}). $K_{p,\rm disk}$ is plotted as the "solar target" curve in Fig. \ref{Kp plot} 
  
  Another "indirect" contribution would be solar wind implantation of $^{10}$Be produced on the protoSun \citep{BrickerCaffee2010}. Although \citet{BrickerCaffee2010} originally envisioned implantation on bare solids, one can equally envision implantation on the gas disk surface (followed by vertical mixing); this would amount to adding $(L_{^{10}\rm Be}/2\pi R)\partial f_{\rm SW}/\partial R$ to the right-hand-side of equation (\ref{transport 10Be}) with $L_{^{10}\rm Be}$ the young Sun's $^{10}$Be (number) production rate and $f_{\rm SW}$ the counterpart of $f$. This would amount to an effective increase of $K_{p,\rm disk}$ of $(\mathrm{d}f_{\rm SW}/\mathrm{d}f) L_{^{10}\rm Be}/(L_{10}c_{\rm O})$. If I set $\mathrm{d}f_{\rm SW}/\mathrm{d}f(L_{^{10}\rm Be}/L_{10})$ equal to the \textit{present-day} ratio of long-term average 1 AU $^{10}$Be \citep{NishiizumiCaffee2001}\footnote{Taking into account the factor of 4 between the mean flux of $^{10}$Be on a randomly oriented surface ($(2.9\pm 1.2)\times 10^{-2}\:\rm m^{-2}s^{-1}$ on the Moon) and its omnidirectional flux \textit{in vacuo}.} and $>$10 MeV proton \citep{Reedy1996} energy omnidirectional fluxes, this evaluates to $6\times 10^{-20}\:\rm s^2/m^2$. From Fig. \ref{Kp plot}, at face value, solar wind implantation thus appears negligible compared to local (in-disk) spallation (except for steep $p\gtrsim 4$) 
but since we do not really know how $^{10}$Be production on the Sun must be extrapolated to its very active early times, the nominal one order-of-magnitude deficit may not warrant definitive conclusions yet.
		
\begin{figure}
\resizebox{\hsize}{!}{
\includegraphics{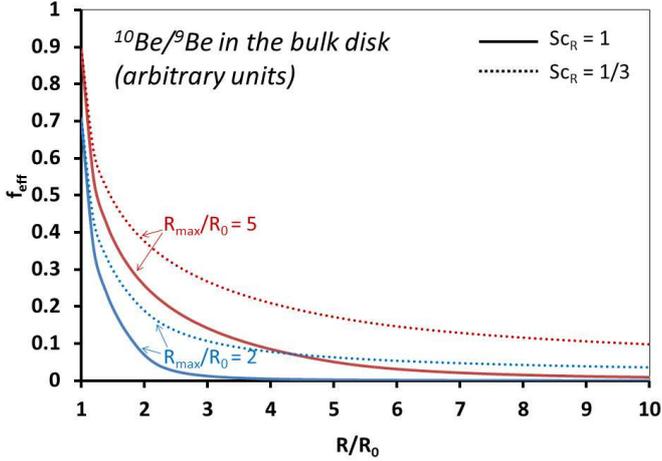}
}
\caption{Plot of $f_{\rm eff}$ (equation (\ref{feff eq})) as a function of heliocentric distances for two values of $R_{\rm max}$ for my toy parameterization of the particle flux distribution (see end of section \ref{production rate}) and two values of the radial Schmidt number, with the lower one (higher relative diffusivity) in dotted lines. $f_{\rm eff}$ is proportional to the $^{10}$Be/$^9$Be ratio (whose absolute magnitude is discussed at equation (\ref{10Be numerical})) in the bulk (gas+solids) disk for given mass accretion rate and proton output and energy distributions.}
\label{feff}
\end{figure}

		Returning to equation (\ref{transport 10Be}), the requirement that $^{10}$Be/$^9$Be vanishes at infinity leads to the following first integration:
		\begin{equation}
		\left(\frac{\partial}{\partial R}-\frac{u_R}{D_R}\right)\left(\frac{^{10}\rm Be}{^9\rm Be}\right)_{\rm disk}=\frac{K_{p, \rm disk}L_{10}}{2\pi R\Sigma D_R}\left(\frac{\rm O}{\rm Be}\right)_{\rm disk}f(R,+\infty)
		\end{equation}
Provided the radial Schmidt number Sc$_R\equiv \nu/D_R$ has a finite lower bound, the requirement that $^{10}$Be/$^9$Be does not diverge at the inner edge of the disk (since $u_R/D_R$ would not be integrable there from equation (\ref{uR})) leads to the unique solution:
\begin{eqnarray}
\label{10Be gas}
\left(\frac{^{10}\rm Be}{^9\rm Be}\right)_{\rm disk}&=& K_{p,\rm disk}L_{10}\left(\frac{\rm O}{\rm Be}\right)_{\rm disk}\Bigg[\frac{f(R,+\infty)}{\dot{M}}\nonumber\\
&&-\int_{R_0}^R\mathrm{exp}\left(\int_{R'}^R \frac{u_R}{D_R}\mathrm{d}R''\right)\frac{\partial}{\partial R}\left(\frac{f(R',+\infty)}{\dot{M}}\right)\mathrm{d}R'\Bigg]\nonumber\\
&=& \frac{K_{p,\rm disk}L_{10}}{\dot{M}}\left(\frac{\rm O}{\rm Be}\right)_{\rm disk}f_{\rm eff},
\end{eqnarray}
where the second equality assumes uniform $\dot{M}$ and Sc$_R$, with
\begin{eqnarray}
\label{feff eq}
f_{\rm eff}\equiv f(R,+\infty) +\int_{R_0}^R \left(\frac{R'^{1/2}-R_0^{1/2}}{R^{1/2}-R_0^{1/2}}\right)^{3\mathrm{Sc}_R}\frac{\partial f(R_0,R')}{\partial R}\mathrm{d}R',
\end{eqnarray}
which is plotted in Fig. \ref{feff}. 

  Interestingly, the result is weakly sensitive to the details of the protoplanetary disk (e.g. turbulence level etc.) with the reference to heliocentric distance being only implicit in $f_{\rm eff}$. This is because the approximate $\propto R^{-2}$ decrease of the flux density is essentially compensated by the $R^2$ factor in the radial transport timescale $t_{\rm vis}$. The above expression can be interpreted as a sum of contributions advected from outer regions (see next subsection, in particular equation (\ref{10Be solid})) and contributions diffused back outward. $^{10}$Be/$^9$Be decreases monotonically outward, falling off roughly as $R^{-3\mathrm{Sc}_R/2}$ outside the irradiated region, conform to the equilibrium gradient of a passive scalar \citep{ClarkePringle1988,JacquetRobert2013}. In the limit Sc$_R \rightarrow 0$, it converges pointwise toward the inner edge value $f(R_0,+\infty)$ and a flat profile, but Sc$_R$ is probably of order unity in MHD turbulence \citep{Johansenetal2006}, not to mention the possibility of laminar wind-driven accretion (and thus even higher Sc$_R$) further out \citep[e.g.][]{Bai2016}. So contrary to earlier statements \citep{Gounelleetal2013,Koopetal2018Ne}, $^{10}$Be production in the gas would entail no spatial uniformity of the $^{10}$Be/$^9$Be ratio.

\subsection{\textit{In situ} $^{10}$Be production in free-floating solids}
\label{Solids}

  I now consider a (sub-)millimeter rocky solid, say a CAI, formed (or more precisely, whose beryllium last equilibrated with the gas) at an heliocentric distance $R_1$. It will certainly inherit the $^{10}$Be/$^9$Be ratio of the local reservoir calculated above, but it should also acquire additional $^{10}$Be so long it wanders through the irradiated region of the disk. The purpose of this subsection is to estimate this contribution. 

  Since the inner disk (where irradiation may occur) is dense, I assume the solid to be tightly coupled to the gas (that is, a gas-grain decoupling parameter $S\ll 1$; \citealt{Jacquetetal2012S}; this is consistent with the lack of bulk chemical fractionation assumed previously). It is also small enough for the thin target approximation \citep[e.g.][]{Sossietal2017} to apply, that is, for equation (\ref{local 10Be}) to apply at the scale of the whole particle as a function of the local (outside) monoenergetic intensities. Since the radial transport timescale $t_{\rm vis}$ is much longer than the vertical transport timescale $H^2/\nu$ with $H=c_s/\Omega_K$ the pressure scale height, I can use the average rate of $^{10}$Be production given by equation (\ref{avg mit Kp}). That is, I view the random vertical motion of the solids due to turbulence as an ergodic process \citep[see e.g. Fig. 8 in][]{Ciesla2010vertical} and identify the time average of this rate for a given solid to the instantaneous average over the long-term probability distribution of same-size solids, which here follows the distribution of the gas (but see appendix \ref{Settling} for the case of finite settling)\footnote{It may also be noted that since the portion of $t_{vis}$ spent within a column $\Sigma_{\rm spall}$ of the surface
	\begin{eqnarray}
	t_{\rm vis}\frac{2\Sigma_{\rm spall}}{\Sigma}&=&\frac{6\pi R^2\Sigma_{\rm spall}}{\dot{M}\left(1-\sqrt{R_0/R}\right)}\nonumber\\
	&=& \frac{5\:\rm a}{1-\sqrt{R_0/R}} \left(\frac{R}{0.5\:\rm AU}\right)^2\left(\frac{10^{-7}\:\rm M_\odot/a}{\dot{M}}\right)\left(\frac{\Sigma_{\rm spall}}{10\:\rm kg/m^2}\right)
	\end{eqnarray}
	is much longer than the weekly periodicity of protostellar flares \citep{Wolketal2005}, the relevant proton luminosity $L_{10}$ in equation (\ref{avg mit Kp}) must be the long-term average luminosity rather than the "characteristic" baseline or the typical flare peak value.}. For this refractory target, the sum on the right-hand-side of equation (\ref{Kp}) is essentially restricted to $k$=O and $i$=$^1$H, $^4$He (with no "indirect reaction" contribution). This corresponds to the "rocky target" curve in Fig. \ref{Kp plot}.
	
\begin{figure}
\resizebox{\hsize}{!}{
\includegraphics{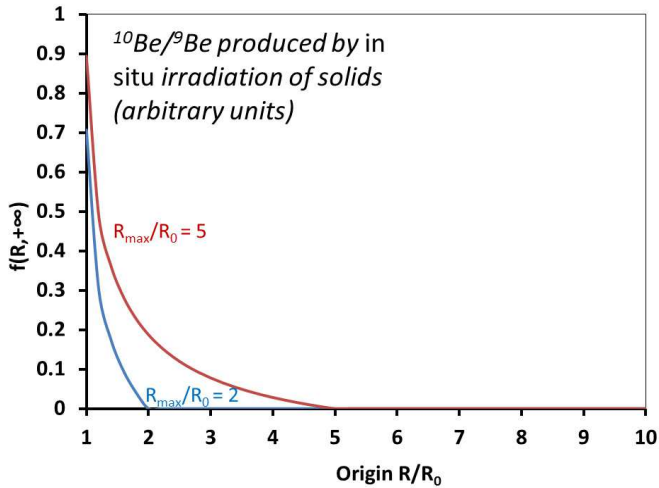}
}
\caption{Plot of $f(R,+\infty)$ as a function of heliocentric distances for two values of $R_{\rm max}$ for my toy parameterization of the particle flux distribution (see end of section \ref{production rate}). This is proportional to the $^{10}$Be/$^9$Be ratio produced by \textit{in} situ irradiation of a refractory solid as a function of its heliocentric distance of origin.}
\label{fsolid}
\end{figure}

  As a result of the turbulent diffusion superimposed on advection by the gas mean velocity, the radial motion of an individual solid has a stochastic character. I will be content to set the "typical timescale" spent crossing a radial bin of width $\Delta R\ll R$ to $|\Delta R/u_R|$. Indeed, even when the flow is opposite to the transport envisioned (as in this subsection), the fraction of a population of solids at $R$ at $t=0$ which has diffused upstream beyond that distance at time $t$ ($\sim\mathrm{erfc}\left((\Delta R-u_Rt)/\sqrt{4D_Rt}\right)/2$, with erfc the complementary error function), is maximum for $t=-\Delta R/u_R$. The total $^{10}$Be produced in the solid during transport between an heliocentric distance $R_1$ and its exit from the irradiated region is then given by:
	\begin{eqnarray}
	\label{10Be solid}
	\left(\frac{^{10}\rm Be}{^9\rm Be}\right)_{\rm CAI} &=& \int_{R_1}^{+\infty}\left\langle\left(\frac{\overset{\bullet}{^{10}\rm Be}}{^9\rm Be}\right)_{\rm tg}\right\rangle_{\rho}\frac{\mathrm{d}R}{|u_R|}\nonumber\\	
	&=& K_{p, \rm CAI}L_{10}\left(\frac{\rm O}{\rm Be}\right)_{\rm CAI}\int_{R_1}^{+\infty} \frac{\partial f(R_0,R)}{\partial R}\frac{\mathrm{d}R}{\dot{M}}\nonumber\\
	&=& \frac{K_{p,\rm CAI}L_{10}}{\dot{M}}\left(\frac{\rm O}{\rm Be}\right)_{\rm CAI}f(R_1,+\infty)
	\end{eqnarray}
 where the last equality assumes a uniform $\dot{M}(R)$. $f(R,+\infty)$ is plotted in Fig. \ref{fsolid}. Unlike the inherited disk value (which benefitted from outward diffusion), nonzero values do not extend to source radii outside the irradiated region. 

\section{Discussion}
\label{Discussion}

\subsection{Magnitude of $^{10}$Be production}
\label{Magnitude}

  Provided a given CAI underwent isotopic equilibration after the last production of $^{10}$Be, the $^{10}$Be/$^9$Be indicated by the slope of an internal isochron is the sum of that inherited from the disk gas and that produced \textit{in situ} in the solid (equations (\ref{10Be gas}) and (\ref{10Be solid}), respectively), which may be written as:
	\begin{equation}
	\label{10Be final}
	\left(\frac{^{10}\rm Be}{^9\rm Be}\right)_{\rm final}=\frac{K_{p,\rm disk}L_{10}f_{\rm sum}}{\dot{M}}\left(\frac{\rm O}{\rm Be}\right)_{\rm disk}
	\end{equation}
	with
	\begin{eqnarray}
	\label{fsum}
	f_{\rm sum}\equiv f(R_1,+\infty)\left(1+\frac{K_{p, \rm CAI}\left(\mathrm{O}/\mathrm{Be}\right)_{\rm CAI}}{K_{p, \rm disk}\left(\mathrm{O}/\mathrm{Be}\right)_{\rm disk}}\right)\nonumber\\
	+\int_{R_0}^{R_1} \left(\frac{R'^{1/2}-R_0^{1/2}}{R_1^{1/2}-R_0^{1/2}}\right)^{3\mathrm{Sc}_R}\frac{\partial f(R_0,R)}{\partial R}\mathrm{d}R'
	\end{eqnarray}
	The contribution of \textit{in situ} production (the term with the "`CAI"' subscripts in the above equation) is subdominant with respect to spallation in the solar gas (that is, $f_{\rm sum}\approx f_{\rm eff}(R_1)$) since, not to mention the ignored incipient settling (see appendix \ref{Settling}), (i) $K_{p,\rm disk}$ outweighs $K_{p,\rm CAI}$ by a factor of $\sim$3 (Fig. \ref{Kp plot}) as it includes additional production channels (e.g. target $^{12}$C, and indirect reactions) and (ii) Be, as a refractory element (half-condensation temperature of 1452 K according to \citet{Lodders2003}), should be concentrated relative to O in a CAI with respect to the overall disk. Indeed Be concentrations typically are of order 0.1-1 ppm in the data of \citet{McKeeganetal2000,Gounelleetal2013}, compared to the CI chondrite value of 25.2 ppb (the latter amounting to $\mathrm{O}/\mathrm{Be}=10^7\approx$ half the solar value; \citealt{Palmeetal2014}). Nevertheless, the fine-grained CAIs of \citet{Sossietal2017} exhibit concentrations down to 6 ppb\footnote{It may nonetheless be wondered whether this average of SIMS points does not underestimate the bulk concentrations given the incompatible behavior of Be in melilite \citep{Paqueetal2014}.}. So, at least for some Be-poor CAIs having spent a random walk in the irradiated region of the disk longer than the "typical timescale" used in subsection \ref{Solids}, \textit{in situ} production may not be entirely negligible. The deviation of the initial boron isotopic composition from solar in some CAIs, too large for an irradiated solar gas, may be a collateral effect of such a contribution \citep{Liuetal2010, Gounelleetal2013}; same for cosmogenic $^3$He and $^{21}$Ne excesses found by \citet{Koopetal2018Ne} in the same type of inclusions. This indicates that, at least for some objects, heliocentric distance $R_1$ was inside the proton-irradiated region of the disk. 
	
	In order to numerically evaluate equation (\ref{10Be final}), I will, as previous authors \citep{Leeetal1998,Gounelleetal2001,Sossietal2017}, use the X-ray luminosity $L_X$ as a proxy for $L_{10}$, since only the former can be measured from distant young stellar objects (although \citet{Ceccarellietal2014} estimated a proton output equivalent to $L_{10}\gtrsim 10^{27}$W, for $L_X<7\times 10^{24}$W, from the ionization level seen toward the single source OMC-2 FIR 4). Since their luminosities are higher than the most powerful flares of the contemporaneous Sun \citep[e.g.][]{Feigelsonetal2002}, solar flares are the least improper analogs in that calibration. \citet{Leeetal1998} derived a scaling $L_{10}=0.09L_X$ by ratioing the total proton and X-ray fluences of the impulsive flares of solar cycle 21 (peaking around 1980). However, while impulsive flares correlate best with X-ray outputs \citep{Leeetal1998}, gradual flares actually dominate the proton fluences at 1 AU \citep{DesaiGiacalone2016}, whatever the ratio at the X point considered by \citet{Leeetal1998} may be. This warrants re-examination of the scaling between X rays and solar energetic particles (SEP) without restricting to impulsive flares. \citet{Emslieetal2012} evaluated the energy outputs of 38 solar eruptive events between 2002 and 2006. Those 21 with reported SEP outputs totaled a SEP output of $1.5\times 10^{25}$J (dominated by $>$10 MeV protons since the energy spectra is shallower below this; \citealt{Mewaldt2006}) and $2.6\times 10^{24}$J in X-rays, hence a ratio of average $L_{10}/L_X\approx 6$ which I adopt as a normalizing value\footnote{Of course, this may shorten the timescales required by \citet{Sossietal2017} at the disk inner edge as well, but they would still represent dozens of orbits.}. Another relevant---if more indirectly for our purpose---dataset is the catalogue of 314 SEP events between 1984 and 2013 of \citet{Papaioannouetal2016}. They represent a total $>$10 MeV proton 1 AU energy fluence of $5.1\times 10^3\:\rm J/m^2$ and total X ray fluence of $54\:\rm J/m^2$ hence $(L_{10}/L_X)\partial f/\partial\mathrm{ln}R/\mathrm{sin}\phi\approx 23$ (which averages different geometries of the flare sources \textit{vis-à-vis} the Earth) assuming isotropic distribution of SEP at 1 AU, suggestive of the same order of magnitude for $L_{10}/L_X$. Equation (\ref{10Be final}) then becomes:
	\begin{eqnarray}
	\label{10Be numerical}
		\left(\frac{^{10}\rm Be}{^9\rm Be}\right)_{\rm final}=6\times 10^{-4}\left(\frac{K_{p,\rm disk}}{9\times 10^{-19}\:\rm s^2/m^2}\right)\left(\frac{\left(\rm O/Be\right)_{\rm disk}}{2.2\times 10^7}\right)\left(\frac{f_{\rm sum}}{0.1}\right)\nonumber\\
\left(\frac{L_{10}/L_X}{6}\right)\left(\frac{L_X}{3\times 10^{23}\: \rm W}\right)\left(\frac{10^{-7}\:\rm M_\odot/a}{\dot{M}}\right)
	\end{eqnarray}
The nominal value complaisantly coincides with the average one measured for regular CAIs in CV chondrites \citep{DavisMcKeegan2014}. This, however, should not hide the considerable uncertainties in several factors, in addition to the $L_{10}/L_X$ discussed above. $K_{p, \rm disk}$, normalized to the value for p=2.5 typical of gradual flares favored by \citet{Sossietal2017}, may lose one order of magnitude or so (depending on the possible implantation 
contributions) for a steeper energy distribution; on the other hand, the normalization value for $f_{\rm sum}$, inspired from the case of ballistic emission
, could be an underestimate if the cosmic rays are focused toward the disk (see Fig. \ref{feff}), but it depends on the essentially unknown magnetic field configuration and turbulent diffusion efficiency. While the predictive value of the model should not thus be overrated, this result does show that, in the current state of the art, the evidence of extinct $^{10}$Be in refractory inclusions \textit{does not} mandate an origin at the very inner edge of the disk, and formation over a wider range of heliocentric distance, say 0.1-1 AU, can be envisioned. This is the main point of this work.

\subsection{Spatio-temporal variations of the $^{10}$Be/$^9$Be ratio}
\label{space-time}

\begin{figure}
\resizebox{\hsize}{!}{
\includegraphics{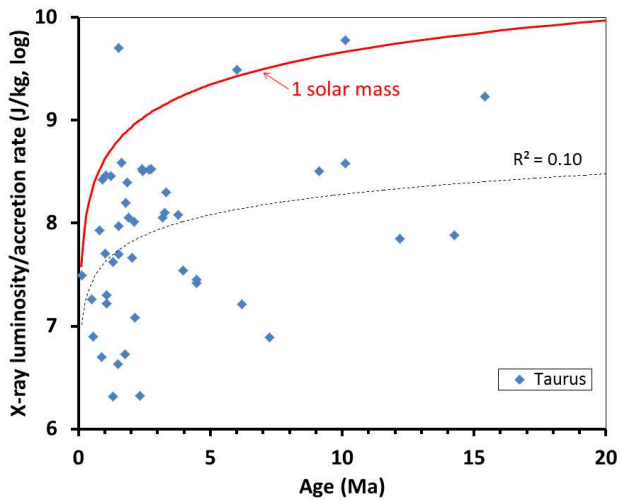}
}
\caption{Plot of $L_X/\dot{M}$ for Taurus as a function of age (data from \citet{Guedeletal2007}). Although a least-square fit (dashed line) shows a systematic increase with age, scatter is evident, in part owing to the diversity of stellar masses. Overplotted is the curve (red, continuous) corresponding to one solar-mass (\citet{Telleschietal2007}; see beginning of section \ref{space-time}). For reference, the normalizations used in equation (\ref{10Be numerical}) amount to $L_X/\dot{M}=10^{7.7}\:\rm J/kg$.}
\label{Lx/Mdot vs age}
\end{figure}

\begin{figure}
\resizebox{\hsize}{!}{
\includegraphics{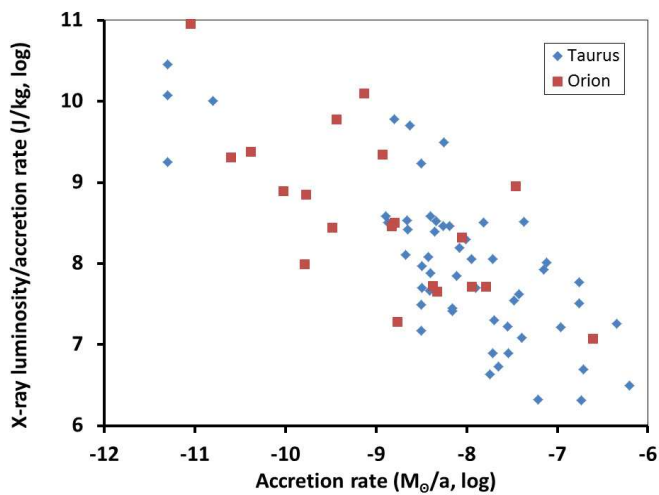}
}
\caption{Plot of $L_X/\dot{M}$ for young stellar objects as a function of $\dot{M}$. Taurus and Orion data are taken from 
\citet{Guedeletal2007} and \citet{Bustamanteetal2016}, respectively (where $L_X/\dot{M}$ also shows little systematic trend with stellar mass).}
\label{Lx vs Mdot}
\end{figure}


  In this model, $^{10}$Be/$^9$Be is proportional to the $L_X/\dot{M}$ which is an observable. Since $L_X$ has only a shallow dependence on time (e.g. $\propto t^{-0.36}$ for one solar mass according to \citealt{Telleschietal2007}), compared to the decrease of the accretion rate ($\propto t^{-1.4}$ according to \citealt{Hartmannetal1998}), $L_X/\dot{M}$ should increase over time (as $t^{1.04}$ if I combine these examples although derived from different data sources; see Fig. \ref{Lx/Mdot vs age}). It may indeed be verified in Fig. \ref{Lx vs Mdot} that $L_X/\dot{M}$ anticorrelates with $\dot{M}$, if with a fair amount of scatter which remind us of the elusive determinants of $L_X$. Since, with decreasing $\dot{M}$, isotherms should recede toward the Sun (see equation (\ref{T})), $f_{\rm sum}$ may also be expected to increase for a given $T(R_1)$ (e.g. a condensation front 
) for a fixed magnetic field configuration. So, barring a strong decrease of $L_{10}/L_X$, $^{10}$Be/$^9$Be should \textit{increase} with time for a given formation and/or equilibration temperature. That is, mass loss from the disk would increase the relative importance of the surficial layers prone to spallation, overcoming the decline in proton luminosity.
	
	This expectation would provide an explanation for the lower $^{10}$Be/$^9$Be of FUN CAIs in CV chondrites \citep{MacPhersonetal2003,Wielandtetal2012} and PLACs in CM chondrites \citep{Liuetal2010} compared to "`regular"' CAIs. Indeed these CAIs are characterized by nucleosynthetic stable isotopic anomalies \citep[e.g. in Ti, Ca;][]{MacPherson2014,Koopetal2018} much in excess of those seen in bulk chondrules \citep{Gerberetal2017} or meteorites \citep{Trinquieretal2009}. This suggests they formed early, perhaps during infall of an isotopically heterogeneous protosolar cloud, before turbulent mixing essentially suppressed the heterogeneities \citep[e.g.][]{Boss2012}
	. Perhaps the $L_X/\dot{M}$ was so low that their $^{10}$Be/$^9$Be was in fact dominated by the background inherited from the protosolar cloud (with perhaps some irradiation from the protostar directly on the envelope before arrival on the disk; \citealt{Ceccarellietal2014}) rather than later-dominating local spallation \citep{Liuetal2010, Wielandtetal2012}.
	
	On the other end of the spectrum, a relatively late formation might explain the high $^{10}$Be/$^9$Be ($7\times 10^{-3}$) of fine-grained CV chondrite CAIs analyzed by \citet{Sossietal2017} or that of Isheyevo CAI 411 ($10^{-2}$; \citealt{Gounelleetal2013}). Indeed the former CAIs have group II rare earth element patterns which are commonly ascribed to condensation in a reservoir previously depleted in an ultrarefractory condensate \citep[e.g.][]{Boynton1989}
	. One may then speculate that this preliminary fractionation took some time to complete (given the tight coupling of millimeter-size solids and gas) so group II CAIs had to be of relatively late formation
. However, while Isheyevo CAI 411 shows $^{26}$Al/$^{27}$Al$< 10^{-6}$---an $^{26}$Al depletion common to many CH chondrite CAIs---, the	fine-grained CV chondrite CAIs tend to exhibit $^{26}$Al/$^{27}$Al ratios no lower than their melted counterparts \citep{MacPhersonetal2012,Kawasakietal2019}, unlike what a later formation would suggest
 in view of the $^{26}$Al half-life of 0.72 Ma \citep[e.g.][]{MacPhersonetal2012}. This conclusion is however predicated on an homogeneous $^{26}$Al distribution which is far from a foregone assumption in early times. In fact, the nucleosynthetic anomaly-bearing CAIs discussed above show reduced $^{26}$Al/$^{27}$Al ($\sim 10^{-5}$; \citealt{MacPhersonetal2014,Parketal2017}) indicating an increase of that ratio in the CAI-forming region between their formation epoch and that of melted CAIs, and perhaps somewhat beyond. Interestingly also, some $^{26}$Al should be produced by spallation along with $^{10}$Be, for their fine-grained CAIs, \citet{Sossietal2017} estimated an increase of $^{26}$Al/$^{27}$Al of $(0.3-1.2)\times 10^{-5}$, comparable to the spread resolved by \citet{MacPhersonetal2012} and \citet{Kawasakietal2019}. At any rate, this calls into question the usefulness of $^{26}$Al as a chronometer during CAI formation, although it may have become more homogenized afterward. Alternatively, the coarse-grained CAIs may have acquired $^{10}$Be/$^9$Be lower than their fine-grained counterparts because of equilibration with their surroundings at higher heliocentric distances, during the localized heating events which melted them. 
	
	This very dependence of $^{10}$Be/$^9$Be on heliocentric distance also implies that the comparatively younger chondrules (or other meteoritic material), which likely formed several AUs away from the Sun (so probably outside the range for \textit{in situ} irradiation), need not have high $^{10}$Be/$^9$Be, not to mention effects of finite time of transport or radioactive decay, as timescales for the chondrule formation epoch (0-3 Ma after CAIs;  \citealt{Nagashimaetal2018, ConnellyBizzarro2018}) or their transport are comparable to the half-life of $^{10}$Be. For a Schmidt number of order unity, one might expect a $2\pm 1$ orders of magnitude depletion with respect to the CAI-forming region (one order of magnitude closer to the Sun), but it is difficult to narrow down. Still, no chondrule $^{10}$Be/$^9$Be values are known yet since silicate crystallization there hardly fractionates Be from B \citep{DavisMcKeegan2014}.

\subsection{Implications on vanadium isotopic systematics}

While it is beyond the scope of the current work to calculate effects of proton irradiation on other isotopic systems, some comments on the reinterpretation of the vanadium-50 excesses (up to 4.4 \textperthousand) reported by \citet{Sossietal2017} are in order. In this framework, spallogenic $^{50}$V would be also dominated by inheritance from the nebula, rather than \textit{in situ} production, but to a lesser extent as the target nuclei/element ratio (essentially Ti/V playing the role of O/Be; \citealt{Sossietal2017}), between two refractory elements, would not markedly differ between a solar gas and a condensate (see second paragraph of subsection \ref{Magnitude}). So the contention by \citet{Koopetal2018Ne} that lack of evidence of \textit{in situ} production of noble gases by solar energetic production in fine-grained CAIs analyzed by \citet{Vogeletal2004}\footnote{which incidentally do show evidence for spallogenic He and Ne, even though it is unclear there is any excess over that due to the recent exposure of the meteoroid to galactic cosmic rays.} prevents a spallogenic origin for their $^{50}$V excesses and favors mass-dependent fractionation upon condensation is not necessarily valid.  Since differential attenuation of the energetic particles modifies their energy distribution at finite penetration columns 
the production ratio (independent of the fluence, and only dependent on $p$) of two nuclides would be affected
. In particular, for a given $p$, since the reactions producing $^{50}$V are efficient at lower energies than for $^{10}$Be \citep{Leeetal1998}, where attenuation is strongest, $^{50}$V production would be comparatively lower. Thus, for given (measured) $^{10}$Be/$^9$Be and $\delta^{51}$V, a steeper slope than in the formalism of \citet{Sossietal2017} would be indicated (for a \textit{given} target composition, which however would have to be revised to solar). Since, if we ratio equation (\ref{avg mit Sigmai}) with its $^{50}$V counterpart, the only difference with the \citet{Sossietal2017} formalism is the presence of $\Sigma_p(E)$ in the integrals in the denominator and the numerator, one may surmise that a shift of $p$ comparable to the power law exponent of $\Sigma_p(E)$ in the CSDA regime (about 1.82; see Appendix \ref{Kp calculation}) would largely cancel out its effect and restore the observed ratio. Although a steeper $p$ would diminish $^{10}$Be production for a given $L_{10}$, this would hardly affect our conclusions given the other uncertainties discussed at the end of section \ref{Magnitude}. This also does not take into account the neutron contribution alluded to in subsection \ref{production rate}, which may alter inferences on the energy distribution. Clearly, a dedicated study on the simultaneous effects expected for different isotopic systems in this scenario, whose proportions are independent of the uncertainties on the fluences, would be worthwhile.

\subsection{X-rays and D/H fractionation}

  As mentioned above, X-ray emission should accompany cosmic ray flares from the Sun. These could also leave isotopic fingerprints (though unrelated to spallation) in meteorites. Indeed \citet{Gavilanetal2017} linked deuterium enrichment of chondritic organic matter (previously ascribed by \citet{Remusatetal2006,Remusatetal2009} to ionizing radiation) to X-ray irradiation near the surface of the disk. The energy fluence $F$ expected for organic matter wandering in the outer disk can be calculated with the same formalism as above if I replace $\sigma_{ki}(E)(i/\mathrm{H})_{\rm CR}(k/\mathrm{Be})_{\rm tg}$ by $E$, so I end this discussion with this short aside. For an attenuation column $\Sigma_X(E)=0.15\:\rm kg/m^2 (E/1\:\rm keV)^n$ with $n=2.485$ and an energy distribution $E I_X\propto \mathrm{exp}\left(-E/(k_BT_X\right))$ with $T_X$ the emission temperature \citep{IgeaGlassgold1999}, the energy-weighted average of the former (the equivalent of $K_p$) is $\Sigma_X(k_BT_X)\Gamma (n+1)$. For distant part of the (flared) disk, I can use $f(R_0,R)=H_X/R$ (with $H_X$ the X-ray absorption height, likely a few times $H$ by analogy with Appendix \ref{Settling}) so that the net fluence has the fairly well-constrained expression (modified after equation (\ref{10Be solid})):
	\begin{eqnarray}
	F &=& \Gamma (n+1)\Sigma_X\left(k_BT_X\right)\frac{L_X}{\dot{M}}\left[\frac{H_X}{R}\right]\nonumber\\
	&=& 2\times 10^7\:\mathrm{J/m^2}\left(\frac{\Gamma (n+1)}{3}\right)\left(\frac{\Sigma_X(1\:\rm keV)}{0.15\:\rm kg/m^2}\right)\left(\frac{k_BT_X}{1\:\rm keV}\right)^n\nonumber\\
&&\left(\frac{L_X}{3\times 10^{23}\:\rm W}\right)\left(\frac{10^{-8}\:\rm M_\odot/a}{\dot{M}}\right)\left(\frac{\left[H_X/R\right]}{0.1}\right),
	\end{eqnarray}
with $\left[H_X/R\right]$ the difference of $H_X/R$ between the locus of formation of the organic matter (or the location inward of which settling ceases to be important, i.e. the $S=1$ line of \citet{Jacquetetal2012S}; see appendix \ref{Settling}) and that of accretion in the chondrite parent body. This is 5 orders of magnitude short of the critical fluence of $5\times 10^{27}\:\rm eV/cm^2=8\times 10^{12}\:\rm J/m^2$ indicated by the experiments of \citet{Gavilanetal2017} for 0.5-1.3 keV photons (and much shorter than \citet{Gavilanetal2017}'s astrophysical estimate as well, which erroneously used X-ray attenuation at surface layers as representative of the bulk). While the steady-state approximation may overestimate the surface density in the outer regions of interest (which nevertheless should not allow efficient settling of the grains), alleviating it would hardly bridge the gap. Nevertheless, the quantitative assessment of isotopic effects of irradiation, whether X rays (in particular at higher energies) or other parts of the spectrum, is still in its infancy and this additional calculation is essentially intended for future applications in similar scenarios.

\section{Conclusion}
\label{Conclusion}

I have analytically investigated the production of short-lived radionuclide beryllium-10 in surface layers of the disk irradiated by protosolar flares. 
 I found that $^{10}$Be production in the gas outweighs $^{10}$Be production in solids after condensation because the gas contains a greater breadth of suitable target nuclides (e.g. $^{12}$C, $^1$H etc.) and incurs less dilution by stable refractory Be. Taking into account incipient settling and possible implantation of solar wind-borne $^{10}$Be would further widen the difference. Although many uncertainties remain on the magnetic field configuration, the scaling of cosmic rays with X-ray luminosities etc., it does appear that this model can reproduce $^{10}$Be/$^9$Be ratios measured in CAIs. Therefore, the past presence of $^{10}$Be does not require (at least at present) that CAIs formed at the inner edge of the disk and allow formation at a fraction of an AU, as thermal models would suggest, and more in line with evidence for a genetic link with their host carbonaceous chondrites (abundance, fraction of the refractory budget; \citealt{Jacquetetal2012S}). If this model holds true, an interesting corollary (barring strong variations of the energetic protons/X-ray ratio) is that the oldest CAIs should have the lowest $^{10}$Be/$^9$Be ratios, which would explain those of nucleosynthetic anomalies-bearing CAIs ((F)UN, PLAC). This would also suggest that the fine-grained group II CAIs in CV chondrites measured by \citet{Sossietal2017} were a relatively late generation of refractory inclusions, a possibility which remains to be explored. This does not mean that chondrules, which formed at comparatively much larger heliocentric distances, should have high $^{10}$Be/$^9$Be, since this ratio decreases outward, following passive diffusion outside the irradiated inner disk. I finally note that the same formalism allows an estimate of fluences of X-rays (produced in the same protosolar flares) or other types of radiations, on aggregates freely floating in the disk which can be e.g. compared to experimental evidence of D/H fractionation in meteoritic organic matter by irradiation.

\section*{Acknowledgments}
I am grateful to the organizers and participants of the workshop "Core to Disk", a program of the $\psi$2 initiative which took place at the Institut d'Astrophysique Spatiale in Orsay from May 14 to June 22, 2018, in particular Patrick Hennebelle and Matthieu Gounelle, who elicited my interest in the subject. I also thank Dr Ming-Chang Liu, Manuel G\"{u}del and Thomas Preibisch who kindly answered my requests for information. Comments by an anonymous reviewer greatly improved the clarity of the result sections as well as the discussion of other isotopic systems. This work is supported by ANR-15-CE31-004-1 (ANR CRADLE).

\bibliographystyle{aa}
\bibliography{bibliography}


\begin{appendix}

\section{Steady-state disk with infall}
\label{Infall}

In this appendix, I study the effects of infall on the steady-state solution for the disk. I adopt the \citet{CassenMoosman1981} expression for the infall source term per unit area $\dot{\Sigma}_{\rm in}$ for a cloud in solid-body rotation, assuming bipolar jets do not suppress it in the inner disk. The mass accretion rate is no longer constant but obeys:
\begin{eqnarray}
-\frac{\partial\dot{M}}{\partial R}&=&2\pi R\dot{\Sigma}_{\rm in}\nonumber\\
&=&\frac{\dot{M_{\rm in}}}{2R_C\sqrt{1-R/R_C}}\theta(R_C-R)
\end{eqnarray}
with $R_C$ the instantaneous centrifugal radius, $\dot{M}_{\rm in}$ the total infall rate on the star+disk system and $\theta$ the Heaviside function. Treating all matter arriving inside $R=R_0$ as directly accreted by the star, this can be integrated as:
\begin{equation}
\dot{M}=\dot{M}_\star - \dot{M}_{\rm in}\left(1-\sqrt{1-\frac{R}{R_C}}\right),
\end{equation}
with $\dot{M}_\star$ the mass accretion rate of the star. If the whole disk can be treated as in steady-state, we should have $\dot{M}_\star=\dot{M}_{\rm in}$ but I refrain from making this identification immediately as the steady-state approximation may only hold locally in the inner regions and/or in the absence of infall ($\dot{M}_{\rm in}=0$) altogether (as assumed in the main text). 

  Angular momentum conservation reads \citep{CassenMoosman1981}:
	\begin{equation}
	\label{angular momentum conservation}
	\dot{M}=6\pi R^{1/2}\frac{\partial}{\partial R}\left(R^{1/2}\Sigma\nu\right)+2\pi R^2\Sigma \frac{\dot{M}_\star}{M_\star}+4\pi R^2\dot{\Sigma}_{\rm in}\left(1-\sqrt{\frac{R}{R_C}}\right),
	\end{equation}
with $M_\star$ the stellar mass. If the latter is much bigger than the mass of the disk, the middle term on the right-hand-side can be neglected upon integration (after division by $R^{1/2}$), which yields:
\begin{eqnarray}
\left(\dot{M}_\star-\dot{M}_{\rm in}\right)\left(1-\sqrt{\frac{R_0}{R}}\right)\nonumber\\+\dot{M}_{\rm in}\left(\frac{R_C}{R}\right)^{1/2}\left[\sqrt{1-x}\left(x^{1/2}-\frac{x+2}{3}\right)\right]^{R/R_C}_{R_0/R_C}=3\pi\Sigma\nu.
\end{eqnarray}
For $R\ll R_C$ this approximates to:
\begin{eqnarray}
3\pi\Sigma\nu &\approx & \left(\dot{M}_\star-\dot{M}_{\rm in}\frac{R+\sqrt{RR_0}+R_0}{2R_C}\right)\left(1-\sqrt{\frac{R_0}{R}}\right)\nonumber\\
& \approx & \dot{M}\left(1-\sqrt{\frac{R_0}{R}}\right)\left(1-\frac{\dot{M}_{\rm in}}{\dot{M}_\star}\frac{\sqrt{RR_0}+R_0}{2R_C}\right)
\end{eqnarray}
which is thus a modest correction to equation (\ref{steady standard}).

  For $R\gg R_0$ and $\dot{M}_\star=\dot{M}_{\rm in}$, I have:
	\begin{eqnarray}
	\dot{M}_{\rm in}\Bigg[\sqrt{1-\frac{R}{R_C}}\Bigg(1 &-& \frac{\left(R/R_C\right)^{1/2}}{3}\Bigg)+\frac{2}{3}\left(\frac{R_C}{R}\right)^{1/2}\left(1-\sqrt{1-\frac{R}{R_C}}\right)\Bigg]\nonumber\\
&=&3\pi\Sigma\nu.
	\end{eqnarray}
	The factor between brackets decreases from 1 to 2/3 when $R$ increases from 0 to $R_C$. Beyond the centrifugal radius, $\dot{M}=0$ and, from equation (\ref{angular momentum conservation}), $R^{1/2}\Sigma\nu$ becomes constant so that:
	\begin{equation}
	3\pi\Sigma\nu=\frac{2}{3}\dot{M}_{\rm in}\left(\frac{R_C}{R}\right)^{1/2}
	\end{equation}
	For bounded $\alpha T$, $2\pi R\Sigma$ is not integrable, but this steady-state solution, which merely sets an upper bound to the disk viscous expansion, can only hold for distances of $t_{\rm vis}(R)$ shorter than the timescale of evolution of the system\footnote{Mathematically, inclusion of the hitherto neglected middle term of the right-hand-side of equation (\ref{angular momentum conservation}) would suffice to make the disk mass finite but on a too large radial scale for it to be relevant}.

\section{Calculation of $K_p$}
\label{Kp calculation}

  I take the attenuation column for protons as the minimum between the continuous slowing-down approximation (CSDA) and diffusion regimes of \citet{Padovanietal2018}, after integrating their equations E.3 and 21\footnote{For the latter, after a change of variable $E_0=E/\mathrm{sin}\theta$, I use the identity:
\begin{equation}
\int_0^{\pi/2}\mathrm{sin}^a\theta\mathrm{d}\theta = \frac{\sqrt{\pi}}{2}\frac{\Gamma\left(\frac{a+1}{2}\right)}{\Gamma\left(\frac{a}{2}+1\right)}
\end{equation}
(for derivation, mutliply the left-hand-side by $\Gamma(1+a/2)=2\int_0^{+\infty} r^{a+1}\mathrm{exp}(-r^2)\mathrm{d}r$ and rewrite the resulting double integral in Cartesian coordinates).
}, respectively: 
\begin{eqnarray}
\Sigma_{p,H}(E) &=& m\mathrm{min}\left[\frac{E^{1+s}}{A\left(p-1\right)},\frac{1}{2(p-1)}\sqrt{\frac{E\left(\alpha_p-1\right)}{3\sigma_{MT}L_p(E)}}\frac{\Gamma\left(\frac{p-1}{\alpha_p-1}+1\right)}{\Gamma\left(\frac{p-1}{\alpha_p-1}+\frac{1}{2}\right)}\right]\nonumber\\
&=& \frac{1}{p-1}\mathrm{min}\Bigg[1.2\: \rm kg/m^2\left(\frac{E}{10\:\rm MeV}\right)^{1.82},\nonumber\\
&& 6\times 10^{2}\:\rm kg/m^2\left(\frac{1\:\rm GeV}{E}\right)^{0.14}\frac{\Gamma\left(\frac{p-1}{\alpha_p-1}+1\right)}{\Gamma\left(\frac{p-1}{\alpha_p-1}+\frac{1}{2}\right)}\Bigg]
\end{eqnarray}
where $A=1.77\times 10^{-6}\:\rm eV^{1+s}m^2$, $s=0.82$, $\alpha_p=1.28$ (noted $\alpha$ in \citet{Padovanietal2018}), $\sigma_{MT}=10^{-30}\:\rm m^2$, $L_p(E)=10^{-21}\:\rm m^2 eV(E/\rm GeV)^{\alpha_p} $ are defined in \citet{Padovanietal2018} and $\Gamma$ is Euler's gamma function. For the other species, I adopt $\Sigma_{p,i}=\Sigma_{p,H}A_i/Z_i^2$, with $Z_i$, $A_i$ its atomic and mass numbers of species $i$ \citep[e.g.][]{Gounelleetal2001}. The transition between the CSDA and the diffusion regime thus occurs at:
\begin{equation}
E_{\rm crit}=0.33\:\rm GeV \left(\frac{\Gamma\left(\frac{p-1}{\alpha_p-1}+1\right)}{\Gamma\left(\frac{p-1}{\alpha_p-1}+\frac{1}{2}\right)}\right)^{0.51}
\end{equation}

  As to the cross sections, similar to \citet{Deschetal2004}, I adopt the \citet{Yanasaketal2001} formulation
		\begin{equation}
		\sigma_{ki}(E)=\sigma_0 \mathrm{min}\left(\left(\frac{E}{E_0}\right)^x,1\right)\theta (E-E_{\rm th})
		\end{equation}
	with $\sigma_0$, $E_{\rm th}$, $E_0$, $x$ constants (for a given $(k,i)$) and $\theta$ is again the Heaviside function. I fitted those for $i=^4$He using the \citet{Langeetal1995} data (unknown to \citet{Yanasaketal2001}), with power-law fitting below 30 MeV/nucleon and the flat regime being assigned the maximum cross section measured (rather than a fit beyond the aforementioned threshold, as this regime is only incipient in the energy range studied). The resulting parameters are presented in table \ref{data}.

\begin{table}
\caption{Cross section parameters}
\label{data}
\begin{tabular}{c c c c c}
\hline \hline
Reaction & $E_{\rm th}$ (MeV) & $\sigma_0$ (mb) & $E_0$ (MeV) & $x$\\
\hline
$^{12}$C(p,x)$^{10}$Be & 27.4 & 3.79 & 1037 & 0.53 \\
$^{14}$N(p,x)$^{10}$Be & 32.1 & 1.75  & 671 & 0.715\\
$^{16}$O(p,x)$^{10}$Be & 34.6 & 2.52 & 798 & 0.962\\
$^{12}$C($\alpha$,x)$^{10}$Be & 19.675 & 9.38 & 34 & 1.4\\
$^{14}$N($\alpha$,x)$^{10}$Be & 11.05 & 5.26 & 30 & 3.4\\
$^{16}$O($\alpha$,x)$^{10}$Be & 13.45 & 4.35 & 29 & 4.3\\
\hline
\hline
\end{tabular}
\end{table}

  Then, I have, for $E_{\rm th}\leq E_0\leq E_{\rm crit}$:
	\begin{eqnarray}
	\int_0^{+\infty}\sigma_{ki}(E)\Sigma_{p,i}E^{-p}\mathrm{d}E = \sigma_0\frac{\Sigma_{p,i}^{\rm CSDA}(E_0)}{E_0^{p-1}}\Bigg(\frac{1-(E_{\rm th}/E_0)^{2+s+x-p}}{2+s+x-p}\nonumber\\
+\frac{(E_{\rm crit}/E_0)^{2+s-p}-1}{2+s-p}
	+\frac{(E_{\rm crit}/E_0)^{2+s-p}}{p+(\alpha_p-3)/2}\Bigg),
	\end{eqnarray}
and for $E_{\rm th}\leq E_{\rm crit}\leq E_0$:
\begin{eqnarray}
	\int_0^{+\infty}\sigma_{ki}(E)\Sigma_{p,i}E^{-p}\mathrm{d}E \nonumber\\= \sigma_0\frac{\Sigma_{p,i}^{\rm CSDA}(E_0)}{E_0^{p-1}}
\Bigg(\frac{(E_{\rm crit}/E_0)^{2+s+x-p}-(E_{\rm th}/E_0)^{2+s+x-p}}{2+s+x-p}\nonumber\\
	+\frac{(E_{\rm crit}/E_0)^{s+(\alpha_p+1)/2}-(E_{\rm crit}/E_0)^{2+s+x-p}}{x-p-(\alpha_p-3)/2}
+\frac{(E_{\rm crit}/E_0)^{s+(\alpha_p+1)/2}}{p+(\alpha_p-3)/2}\Bigg),
	\end{eqnarray}
	whose contributions can be summed to yield $K_p$ (equation (\ref{Kp})).

\section{Height of spallation layer and effect of settling on $^{10}$Be production in solids}
\label{Settling}

The spallation layer column $\Sigma_{\rm spall}$ may be defined as:
	\begin{equation}
	\Sigma_{\rm spall}\equiv \frac{1}{\left(\overset{\bullet}{^{10}\rm Be}/^9\rm Be\right)_{\rm tg, +}(0)}\int_0^{+\infty}\left(\frac{\overset{\bullet}{^{10}\rm Be}}{^9\rm Be}\right)_{\rm tg,+}\mathrm{d}\Sigma'
=\langle\Sigma_p(E)\rangle_{\sigma E^{-p}} \frac{\mathrm{sin}\phi}{2}
\end{equation}
Assuming a vertically isothermal disk (and thus a gaussian gas density $\rho(z)=\Sigma\mathrm{exp}(-(z/H)^2/2)/(\sqrt{2\pi}H)$), the height $H_{\rm spall}$ of this layer obeys:
\begin{equation}
\Sigma_{\rm spall}=\frac{\Sigma}{2}\mathrm{erfc}\left(\frac{H_{\rm spall}}{H\sqrt{2}}\right).
\end{equation}
For $\Sigma\sim 10^{5\pm 1}\:\rm kg/m^2$ and $\Sigma_{\rm spall}=10^{1\pm 1}\:\mathrm{kg/m^2}$, I have $\chi\equiv H_{\rm spall}/H \approx 3\pm 1$, so that:
\begin{equation}
\frac{H_{\rm spall}}{R}=0.16\left(\frac{\chi}{3}\right)\left(\frac{R}{0.5\:\rm AU}\right)^{1/2}\left(\frac{T}{1500\:\rm K}\right)^{1/2}.
\end{equation}

Since, for $x\gtrsim 2$, I have\footnote{This is because erfc$'(x)=-2\mathrm{exp}(-x^2)/(\sqrt{\pi})\sim -\mathrm{exp}(-x^2)(2+1/x^2)/\sqrt{\pi}=(\mathrm{exp}(-x^2)/(\sqrt{\pi}x))'$.}
\begin{equation}
\label{erfc}
\mathrm{erfc}(x)\approx \frac{\mathrm{exp}(-x^2)}{\sqrt{\pi}x},
\end{equation}
the density at $z=H_{\rm spall}$ may be approximated as:
\begin{eqnarray}
\label{rho_spall}
\rho(H_{\rm spall})=\frac{\chi\Sigma_{\rm spall}}{H}=8\times 10^{-9}\:\mathrm{kg/m^3}\left(\frac{\chi}{3}\right)\left(\frac{\Sigma_{\rm spall}}{10\:\rm kg/m^2}\right)\left(\frac{1500\:\rm K}{T}\right)^{1/2}\nonumber\\
\left(\frac{0.5\:\rm AU}{R}\right)^{3/2}.
\end{eqnarray}
The nominal value (corresponding to a number density of $2\times 10^{12}\:\rm cm^{-3}$) is half-way in the range studied by \citet{UmebayashiNakano1981} so the neutron contribution (ignored in this paper) can be seen in their Fig. 2 and 5 to be limited. Then, the settling parameter \citep{Jacquetetal2012S} for a solid of density $\rho_s$ and radius $a$ there is:
\begin{eqnarray}
S_z(H_{\rm spall})&=&\sqrt{\frac{\pi}{8}}\frac{\rho_s a}{\rho(H_{\rm spall})c_s\delta_z}\nonumber\\
&\approx & \sqrt{\frac{\pi}{8}}\frac{\rho_sa}{\Sigma_{\rm spall}\delta_z\chi}\nonumber\\
&=& 0.2\left(\frac{\rho_s a}{1\:\rm kg/m^2}\right)\left(\frac{10\:\rm kg/m^2}{\Sigma_{\rm spall}}\right)\left(\frac{10^{-1}}{\delta_z}\right)\left(\frac{3}{\chi}\right)
\end{eqnarray}
with $\delta_z$ the vertical diffusion coefficient normalized to $c_s^2/\Omega_K$ (which should be of order $\alpha$). 

  Simulations of ideal MHD turbulence indicate $\delta_z\sim\alpha\propto\rho^{-1}$ over a large extent of the disk thickness, although it should stall in the corona \citep[e.g.][and references therein]{Jacquet2013}. Since $S_z$ may be expected to be $\lesssim 1$ there from the above evaluation, it may not be a dramatic underestimate to treat it as constant throughout the vertical thickness of the disk. Under that assumption, a population of identical solids has a density \citep{Jacquet2013}
	\begin{equation}
	\rho_c(z)=\frac{\Sigma_c\sqrt{1+S_z}}{\sqrt{2\pi}H}\mathrm{exp}\left(-\frac{S_z+1}{2}\left(\frac{z}{H}\right)^2\right)
	\end{equation}
	with $\Sigma_c$ the surface density of the population. The $^{10}$Be production rate averaged over this distribution is then
	\begin{equation}
	\left\langle\left(\frac{\overset{\bullet}{^{10}\rm Be}}{^9\rm Be}\right)_{\rm tg}\right\rangle_{\rho_c}\equiv \frac{2}{\Sigma_c}\int_0^{+\infty}\mathrm{d}\Sigma_c'\left(\frac{\overset{\bullet}{^{10}\rm Be}}{^9\rm Be}\right)_{\rm tg,+}\left(\Sigma'(\Sigma_p')\right)
\end{equation}
	with $\Sigma_c'$ and $\Sigma'$ the columns of the solids and gas, respectively, integrated vertically from the surface. If I let
	\begin{equation}
	\left(\frac{\overset{\bullet}{^{10}\rm Be}}{^9\rm Be}\right)_{\rm tg,+}\equiv \left(\frac{\overset{\bullet}{^{10}\rm Be}}{^9\rm Be}\right)_{\rm tg,+}(0)  g\left(\frac{\Sigma'}{\Sigma_{\rm spall}}\right),
	\end{equation}
	and since
	\begin{equation}
	\Sigma'=\frac{\Sigma}{2}\mathrm{erfc}\left(\frac{\mathrm{erfc}^{-1}\left(2\Sigma_c'/\Sigma_c\right)}{\sqrt{1+S_z}}\right),
	\end{equation}
	this becomes:
	\begin{eqnarray}
	\left\langle\left(\frac{\overset{\bullet}{^{10}\rm Be}}{^9\rm Be}\right)_{\rm tg}\right\rangle_{\rho_c}&=&\frac{\Sigma}{2\Sigma_{\rm spall}}\int_0^{+\infty} g\left(\frac{\Sigma}{2\Sigma_{\rm spall}}\mathrm{erfc}\left(\frac{\mathrm{erfc}^{-1}(x)}{\sqrt{1+S_z}}\right)\right)\mathrm{d}x \left\langle\left(\frac{\overset{\bullet}{^{10}\rm Be}}{^9\rm Be}\right)_{\rm tg}\right\rangle_{\rho}\nonumber\\
	& \approx &  \frac{\Sigma}{2\Sigma_{\rm spall}}\int_0^{+\infty} g\left(\frac{\Sigma}{2\Sigma_{\rm spall}}
	\frac{x^{1/(S_z+1)}\sqrt{S_z+1}}{\left(\sqrt{\pi}\mathrm{erfc}^{-1}(x)\right)^{S_z/(S_z+1)}}
	\right)\mathrm{d}x \nonumber\\
&& \left\langle\left(\frac{\overset{\bullet}{^{10}\rm Be}}{^9\rm Be}\right)_{\rm tg}\right\rangle_{\rho}
	\end{eqnarray}
where I have used approximation (\ref{erfc}),with the corollary $\mathrm{erfc}^{-1}(z)\approx \sqrt{-\mathrm{ln}(\sqrt{\pi}z\mathrm{erfc}^{-1}(z))}$ for $z\ll 1$. Roughly speaking, the integrand only contributes ($\sim 1$) for a $g$ argument $\lesssim 1$, that is $x\lesssim (\sqrt{\pi}\mathrm{erfc}^{-1}(x))^{S_z}(2\Sigma_{\rm spall}/\Sigma\sqrt{S_z+1})^{S_z+1}$ so that:
\begin{equation}
\left\langle\left(\frac{\overset{\bullet}{^{10}\rm Be}}{^9\rm Be}\right)_{\rm tg}\right\rangle_{\rho_c}\sim \frac{1}{\sqrt{S_z+1}}\left(\sqrt{\pi\mathrm{ln}(\Sigma/\Sigma_{\rm spall})}\frac{2\Sigma_{\rm spall}}{\Sigma}\right)^{S_z}\left\langle\left(\frac{\overset{\bullet}{^{10}\rm Be}}{^9\rm Be}\right)_{\rm tg}\right\rangle_{\rho}.
\end{equation}
So $^{10}$Be production should quickly drop as $S_z$ approaches unity 
 and become negligible for $S_z\gtrsim 1$.

\end{appendix}

\end{document}